\documentstyle [12pt, epsfig]{article}
\begin{document}

\hfill {CUMQ/HEP 120}

\hfill {\today}

\vskip 0.5in   \baselineskip 24pt

{
\Large
      \bigskip
      \centerline{ {\Large $b \rightarrow s \gamma$ in the left-right
supersymmetric
    model} }
    }

\vskip .6in
\def\bar{\overline}

\centerline{Mariana Frank \footnote{Email: mfrank@vax2.concordia.ca}
and Shuquan Nie \footnote{Email: sxnie@alcor.concordia.ca}}
\bigskip
\centerline {\it Department of Physics, Concordia University, 1455 De
Maisonneuve Blvd. W.}
\centerline {\it Montreal, Quebec, Canada, H3G 1M8}

\vskip 0.5in

{\narrower\narrower  The rare decay $b \rightarrow s \gamma$ is studied
in the left-right supersymmetric
model. We give explicit expressions for all the amplitudes
associated with the supersymmetric contributions
coming from gluinos, charginos and neutralinos in the model to
one-loop level. The branching ratio is enhanced
significantly compared to the standard model and minimal
supersymmetric standard model values by contributions from the
right-handed gaugino and squark sector. We give numerical results
coming from the leading order contributions. If the only
source of flavor violation comes from the CKM matrix, we constrain
the scalar fermion-gaugino sector. If
intergenerational mixings are allowed in the squark mass matrix, we
constrain such supersymmetric sources of flavor
violation. The decay $b \rightarrow s \gamma$ sets constraints on the
parameters of the model
and provides distinguishing signs from other supersymmetric scenarios. }

PACS number(s): 12.38.Bx, 12.60.Jv, 12.35.Hw

\newpage

\section{Introduction}

The experimental and theoretical investigation of the inclusive decay
$B \rightarrow X_s \gamma$ is an
important benchmark for physics beyond the Standard Model (SM).
Experimentally, the inclusive decay $B \rightarrow X_s
\gamma$ has been measured at ALEPH \cite{aleph}, BELLE \cite{belle}
and CLEO \cite{cleo},with BABAR
measurements keenly awaited, giving the following weighted average:
\begin{equation}
BR(B \rightarrow X_s \gamma)=(3.23 \pm 0.41) \times 10^{-4}.
\end{equation}
This present experimental average is in good agreement with the
next-to-leading order predictions in the SM \cite{smbsg}:
\begin{equation}
BR(B \rightarrow X_s \gamma)_{SM}=(3.35 \pm 0.30) \times 10^{-4},
\end{equation}
But this value still allows a large acceptable range for the
inclusive decay \cite{range}:
\begin{equation}
2 \times 10^{-4} < BR(B \rightarrow X_s \gamma) < 4.5 \times 10^{-4}.
\end{equation}
Flavor changing neutral currents (FCNC) are forbidden at the tree level
in the SM. The first SM contributions to the process $b
\rightarrow s
\gamma$ appear at one-loop level through the
Cabibbo-Kobayashi-Maskawa (CKM) flavor
mixing. Despite small uncertainties in the theoretical
evaluation of the branching ratio, agreement between experiment and
theory is impressive, and this fact is used to
set constraints on the parameters for beyond the SM
scenarios, such as two-Higgs-Doublet-Models (2HDM) \cite{2HDM},
left-right symmetric models (LRM) \cite{LR}, and
minimal supersymmetric standard model (MSSM) \cite{SUSY}. Although
attempts have been
made to reconcile $b \rightarrow s \gamma$ with right-handed $b$-quark
decays
\cite{gronau}, a complete analysis for a fully left-right supersymmetric
model is still lacking.

The Left-Right Supersymmetric (LRSUSY) model is perhaps the most natural
extension of the MSSM \cite{history, mohapatra, frank1, huitu}.
Left-right supersymmetry is based on the group $SU(2)_L
\times SU(2)_R \times U(1)_{B-L}$, which would then break
spontaneously to $SU(2)_L \times U(1)_Y$ \cite{history}. LRSUSY was
originally seen as a natural way
to suppress rapid proton decay
\cite {frank1} and has recently received renewed attention for providing
small neutrino masses and lepton radiative decays \cite{leptonsusylr}.
Besides being a
plausible symmetry itself, LRSUSY models have the added attractive
features that
they can be embedded in a supersymmetric grand unified theory such as
$SO(10)$ \cite{SO10}. Another support for left-right theories is
provided by
building realistic brane worlds from Type I strings. This involves
left-right
supersymmetry, with supersymmetry broken either at the string scale
$M_{SUSY} \approx 10^{10-12}$ GeV, or at $M_{SUSY} \approx 1$ TeV, the
difference having implications for gauge unification \cite{string}.

In this paper we study all contributions of the LRSUSY model to the
branching ratio of $b \rightarrow s \gamma$ at one-loop level. The decay $b
\rightarrow s \gamma$ can be mediated by left-handed and right-handed
W bosons and charged Higgs bosons as
in nonsupersymmetric case, but also by
charginos, neutralinos and gluinos. The structure of the LRSUSY
provides a significant contributions
to the decay $b \rightarrow s \gamma$ from the right-handed squarks
and an enlarged
gaugino-Higgsino sector with right-handed couplings, which is not as
constrained as the
right-handed gauge sector in left-right symmetric models. We anticipate
that these would contribute a large enhancement of the decay rate
and would constrain some of the parameters of the model.

The paper is organized as follows. We describe the structure of the model
in Sec. II, with
particular emphasis on the gaugino-Higgsino and squark structure. In
Sec. III, we give the
supersymmetric contributions in LRSUSY to the decay
$b
\rightarrow s
\gamma$. We confront the calculation with
experimental results in Sec. IV, where we present the numerical analysis to
constrain the
parameters of the model for two scenarios: one with CKM flavor mixing
only, the other
including supersymmetric soft breaking flavor violation. We reach our
conclusions in Sec. V.

\section{The Model}

The LRSUSY electroweak symmetry group, $SU(2)_{L}\times 
SU(2)_{R}\times U(1)_{B-L}$,
has matter
doublets for both left- and right-handed fermions and their
corresponding left-
and right-handed scalar partners (sleptons and squarks)~\cite{frank1}.
In the gauge sector,
corresponding to $SU(2)_{L}$ and $SU(2)_{R}$, there are triplet
gauge bosons $(W^{+}, W^{-},W^{0})_{L}$, $(W^{+}, W^{-},W^{0})_{R}$,
respectively, and a singlet
gauge
boson $V$ corresponding to $U(1)_{B-L}$, together with their
superpartners.
The Higgs sector of this model
consists of two Higgs bi-doublets, $\Phi_{u}(\frac{1}{2},\frac{1}{2},0)$
and
$\Phi_{d}(\frac{1}{2},\frac{1}{2},0)$, which are required to give masses
to
the up and down quarks.  The spontaneous symmetry breaking of the group
$SU(2)_{R}\times U(1)_{B-L}$ to the hypercharge symmetry group
$U(1)_{Y}$ is
accomplished by giving vacuum expectation values to a pair of Higgs
triplet fields
$\Delta_{L}(1,0,2)$  and $\Delta_{R}(0,1,2)$, which transform as the
adjoint
representation of $SU(2)_R$. The choice of two triplets (versus four
doublets) is
preferred because with this choice a large Majorana mass can be
generated (through
the see-saw mechanism) for the right-handed neutrino and a small one for
the left-handed neutrino~\cite{mohapatra}.
In addition to the triplets $\Delta_{L,R}$, the model must contain two
additional triplets, $\delta_{L}(1,0,-2)$ and $\delta_{R}(0,1,-2)$, with
quantum number $B-L= -2$, to insure cancellation of the anomalies which
would
otherwise occur in the fermionic sector.
The superpotential for the LRSUSY model is:
\begin{eqnarray}
\label{superpotential}
W_{LRSUSY} & = & {\bf h}_{q}^{(i)} Q^T\tau_{2}\Phi_{i} \tau_{2}Q^{c} + {\bf
h}_{l}^{(i)}
L^T\tau_{2}\Phi_{i} \tau_{2}L^{c} + i({\bf h}_{LR}L^T\tau_{2} \Delta_L L
+ {\bf
h}_{LR}L^{cT}\tau_{2}
\Delta_R L^{c}) \nonumber \\
& & + M_{LR}\left [Tr (\Delta_L  \delta_L +\Delta_R
\delta_R)\right] + \mu_{ij}Tr(\tau_{2}\Phi^{T}_{i} \tau_{2} \Phi_{j})
+W_{NR}
\end{eqnarray}
where $W_{NR}$ denotes (possible) non-renormalizable terms arising
from higher scale
physics or Planck scale effects~\cite{recmohapatra}. The presence of
these terms
insures that, when the SUSY breaking scale is above $M_{W_{R}}$, the
ground state is R-parity conserving ~\cite{km}.

The neutral Higgs fields acquire non-zero vacuum
expectation values $(VEV's)$ through spontaneous symmetry breaking:
\begin{eqnarray}
\langle \Delta \rangle_{L,R} = \left(\begin{array}{cc}
0&0\\v_{L,R}&0
\end{array}\right),
~\rm{and}~
\langle \Phi \rangle_{u,d} = \left (\begin{array}{cc}
\kappa_{u,d}&0\\0&\kappa^{\prime}_{u,d} e^{i\omega}
\end{array}\right).
\nonumber
\end{eqnarray}
$\langle \Phi \rangle$ causes the mixing of $W_{L}$ and $W_{R}$
bosons with $CP$-violating
phase $\omega$. The non-zero Higgs $VEV's$
breaks both parity and $SU(2)_{R}$.
In the first stage of breaking, the right-handed gauge bosons, $W_{R}$ and
$Z_{R}$ acquire masses proportional to $v_{R}$ and become much heavier
than the SM (left-handed) gauge bosons $W_{L}$ and $Z_{L}$, which pick
up masses
proportional to $\kappa_{u}$ and $\kappa_{d}$ at the second stage of
breaking.

In the supersymmetric sector of the model there are six singly-charged
charginos, corresponding to $\tilde\lambda_{L}$,
$\tilde\lambda_{R}$, $\tilde\phi_{u}$,
$\tilde\phi_{d}$, $\tilde\Delta_{L}^{\pm}$, and
$\tilde\Delta_{R}^{\pm}$.
The model also has eleven neutralinos, corresponding to
$\tilde\lambda_{Z}$,
$\tilde\lambda_{Z^{\prime}}$,
$\tilde\lambda_{V}$,  $\tilde\phi_{1u}^0$, $\tilde\phi_{2u}^0$,
$\tilde\phi_{1d}^0$,  $\tilde\phi_{2d}^0$, $\tilde\Delta_{L}^0$,
$\tilde\Delta_{R}^0$,  $\tilde\delta_{L}^0$, and
$\tilde\delta_{R}^0$. Although $\Delta_{L}$
is not necessary for symmetry breaking~\cite{huitu}, and is
introduced only for preserving left-right symmetry, both
$\Delta_{L}^{--}({\tilde \Delta_{L}^{--}})$ and its
right-handed counterparts $\Delta_{R}^{--}({\tilde \Delta_{R}^{--}})$ 
play very important
roles in lepton phenomenology of the LRSUSY model. The doubly charged Higgs
and Higgsinos do not
affect quark phenomenology, but the neutral and singly charged
components do, through
mixings in the chargino and neutralino mass matrices. We include only
the
${\tilde \Delta}_R$ contribution in the numerical analysis.

The supersymmetric sources of flavor violation in the LRSUSY model
come from either the
Yukawa potential or the trilinear scalar coupling.

The interaction of fermions with scalar (Higgs) fields has the following
form:
\begin{eqnarray}
\label{eq:yukawa}
{\cal L}_Y= {\bf h}_u\bar{Q}_L \Phi_u Q_R + {\bf h}_d \bar{Q}_L \Phi_d
Q_R\,
+{\bf h}_\nu\bar{L}_L  \Phi_u L_R + {\bf h}_e \bar{L}_L \Phi_d
L_R+\,H.c.;\nonumber \\
{\cal L}_M=i{\bf h}_{LR}(L_L^TC^{-1}\tau_2\Delta_LL_L+
L_R^TC^{-1}\tau_2\Delta_RL_R) + H.c.
\end{eqnarray}
where ${\bf h}_u$, ${\bf h}_d$, ${\bf h}_{\nu}$ and ${\bf h}_e$ are
the Yukawa couplings
for the up and down quarks and neutrino and electron, respectively,
and ${\bf h}_{LR}$ is
the coupling for the triplet Higgs bosons.  LR symmetry requires all
${\bf h}$-matrices to be Hermitean in generation space  and
${\bf h}_{LR}$ matrix to be symmetric.
We present below the gaugino-Higgsino as well as the sfermion structure
of the model, before proceeding with calculation of the branching ratio
of $b
\rightarrow s \gamma$.

\subsection{Charginos}

The terms relevant to the masses of charginos in the Lagrangian are:
\begin{equation}
{\cal L}_C=-\frac{1}{2}(\psi^+, \psi^-) \left ( \begin{array}{cc}
                                                     0 & X^T \\
                                                     X & 0
                                                   \end{array}
                                           \right ) \left (
\begin{array}{c}
                                                            \psi^+ \\
                                                            \psi^-
                                                            \end{array}
                                                     \right ) + H.c. \ ,
\end{equation}
where $\psi^+=(-i \lambda^+_L, -i \lambda^+_R, \tilde{\phi}_{u1}^+,
\tilde{\phi}_{d1}^+, \tilde{\Delta}_R^+)^T$
and $\psi^-=(-i \lambda^-_L, -i \lambda^-_R, \tilde{\phi}_{u2}^-,
\tilde{\phi}_{d2}^-, \tilde{\delta}_R^-)^T$, and:
\begin{equation}
X=\left( \begin{array}{ccccc}
                         M_L & 0 & g_L \kappa_u & 0 & 0 \\
                         0 & M_R & g_R \kappa_u & 0 & 0 \\
                         0 & 0 & 0 & -\mu & 0 \\
                         g_L \kappa_d & g_R \kappa_d & -\mu & 0 & 0 \\
                         0 & \sqrt{2} g_R v_R & 0 & 0 & -\mu
            \end{array}
      \right )
\end{equation}
where we have taken, for simplification, $\mu_{ij}=\mu$. The chargino mass
eigenstates $\chi_i$ are obtained by:
\begin{eqnarray}
\chi_i^+=V_{ij}\psi_j^+, \ \chi_i^-=U_{ij}\psi_j^-, \ i,j=1, \ldots 5,
\end{eqnarray}
with $V$ and $U$ unitary matrices satisfying:
\begin{equation}
U^* X V^{-1} = M_D,
\end{equation}
where $M_D$ is a diagonal matrix with non-negative entries. Positive
square roots of the eigenvalues of
$X^{\dagger} X$ ($X X^{\dagger}$) will be the diagonal entries of
$M_D$ such that:
\begin{equation}
V X^{\dagger} X V^{-1}=U^* X X^{\dagger} (U^*)^{-1}=M_D^2.
\label{equationC}
\end{equation}
The diagonalizing matrices $U^*$ and $V$ are obtained by
computing the eigenvectors corresponding
to the eigenvalues of $X^{\dagger} X$ and $X X^{\dagger}$, respectively.

\subsection{Neutralinos}

The terms relevant to the masses of neutralinos in the Lagrangian are:
\begin{equation}
{\cal L}_N=-\frac{1}{2} {\psi^0}^T Y \psi^0  + H.c. \ ,
\end{equation}
where $\psi^0=(-i \lambda_L^3, -i \lambda_R^3, -i \lambda_V,
\tilde{\phi}_{u1}^0, \tilde{\phi}^0_{u2},
\tilde{\phi}_{d1}^0, \tilde{\phi}^0_{d2}, \tilde{\Delta}_R^0,
\tilde{\delta}_R^0 )^T $, and:
and
\begin{equation}
Y=\left( \begin{array}{ccccccccc}
            M_L & 0 & 0 & \frac{g_L \kappa_u}{\sqrt{2}} & 0 & 0 & -
\frac{g_L \kappa_d}{\sqrt{2}} & 0 & 0 \\
            0 & M_R & 0 & \frac{g_R \kappa_u}{\sqrt{2}} & 0 & 0 &
-\frac{g_R \kappa_d}{\sqrt{2}} & -\sqrt{2}g_R v_R & 0 \\
            0 & 0 & M_V & 0 & 0 & 0 & 0 & 2 \sqrt{2} g_V v_R & 0 \\
            \frac{g_L \kappa_u}{\sqrt{2}} & \frac{g_R \kappa_u}{\sqrt{2}}
&
0 & 0 & 0 & 0 & -\mu & 0 & 0  \\
            0 & 0 & 0 & 0 & 0 & -\mu & 0 & 0 & 0 \\
            0 & 0 & 0 & 0 & -\mu & 0 & 0 & 0 & 0 \\
            -\frac{g_L \kappa_d}{\sqrt{2}} & -\frac{g_R
\kappa_d}{\sqrt{2}}
& 0 & -\mu & 0 & 0 & 0 & 0 & 0  \\
            0 & -\sqrt{2}g_R v_R & \sqrt{2}g_V v_R & 0 & 0 & 0 & 0 & 0 &
-\mu \\
            0 & 0 & 0 & 0 & 0 & 0 & 0 & -\mu & 0
            \end{array}
      \right ).
\end{equation}
The mass eigenstates are defined by:
\begin{equation}
\chi^0_i=N_{ij} \psi^0_j \ (i,j=1,2, \ldots 9),
\end{equation}
where $N$ is a unitary matrix chosen such that:
\begin{equation}
N^* Y N^{-1} = N_D,
\label{equationN}
\end{equation}
and $N_D$ is a diagonal matrix with non-negative entries.
To determine $N$, we take the square of Eq. (\ref{equationN})
obtaining:
\begin{equation}
N Y^{\dagger} Y N^{-1} = N_D^2,
\end{equation}
which is similar to Eq. (\ref{equationC}).

We found it convenient to define the neutralino states in terms of
the photino and
left and right zino states:
\begin{equation}
{\psi^0}^{\prime}=(-i \lambda_{\gamma}, -i \lambda_{Z_L}, -i
\lambda_{Z_R}, \tilde{\phi}_{u1}^0, \tilde{\phi}^0_{u2},
\tilde{\phi}_{d1}^0, \tilde{\phi}^0_{d2}, \tilde{\Delta}_R^0,
\tilde{\delta}_R^0 )^T,
\end{equation}
with:
\begin{eqnarray}
\lambda_{\gamma} &=& \lambda_L^3 \sin \theta_W + \lambda_R^3 \sin
\theta_W + \lambda_V \sqrt{\cos 2 \theta_W} \nonumber \\
\lambda_{Z_L} &=& \lambda_L^3 \cos \theta_W - \lambda_R^3 \sin \theta_W
\tan \theta_W - \lambda_V \sqrt{\cos 2 \theta_W} \tan \theta_W \nonumber
\\
\lambda_{Z_R} &=& \lambda_R^3 \frac{\sqrt{\cos 2 \theta_W}}{\cos
\theta_W}  - \lambda_V \tan \theta_W
\end{eqnarray}
Then the mass matrix $Y$ would be replaced by a matrix $Y^{\prime}$
found to
be:

\begin{equation}
{\small
Y^{\prime}=\left( \begin{array}{ccccccccc}
            m_{\tilde{\gamma}} & 0 & 0 & F  & 0 & 0 &  -F & -2 \sqrt{2} e
v_R & 0 \\
            0 & m_{\tilde{Z}_L} & 0 &  A_1 & 0 & 0 & A_2 & D & 0 \\
            0 & 0 & m_{\tilde{Z}_R} & E   & 0 & 0 & C   &  B & 0 \\
            F  &  A_1 & E & 0 & 0 & 0 & -\mu & 0 & 0  \\
            0 & 0 & 0 & 0 & 0 & -\mu & 0 & 0 & 0 \\
            0 & 0 & 0 & 0 & -\mu & 0 & 0 & 0 & 0 \\
            -F  &  A_2 & C & -\mu & 0 & 0 & 0 & 0 & 0  \\
            -2 \sqrt{2} e v_R  &  D  & B & 0 & 0 & 0 & 0 & 0 & -\mu \\
            0 & 0 & 0 & 0 & 0 & 0 & 0 & -\mu & 0
            \end{array}
      \right ).
}
\end{equation}
with:
\begin{eqnarray}
    m_{\tilde{\gamma}}&=& M_V \cos 2 \theta_W + (M_L+M_R) \sin^2 \theta_W ,
\nonumber \\
    m_{\tilde{Z}_L}&=&M_V \cos 2 \theta_W \tan^2 \theta_W +M_L \cos^2
\theta_W + M_R \sin^2 \theta_W \tan^2 \theta_W , \nonumber \\
    m_{\tilde{Z}_R}&=&M_V \tan^2 \theta_W  + M_R (1-\tan^2 \theta_W) ,
\nonumber \\
    A_1&=& \frac{g}{\sqrt{2} \cos \theta_W} ( \kappa_u \cos^2 \theta_W
-\kappa_d \sin \theta_W) , \nonumber \\
    A_2&=& \frac{g}{\sqrt{2} \cos \theta_W} ( \kappa_d \cos^2 \theta_W
-\kappa_u \sin \theta_W) , \nonumber \\
    B&=&-\sqrt{2} g \frac{1-2 \tan^2 \theta_W}{\sqrt{1-\tan^2 \theta_W}}
v_R , \nonumber \\
    C&=&-\frac{g \sqrt{\cos 2 \theta_W }}{\sqrt{2}\cos \theta_W}
\kappa_u,
\nonumber \\
    D&=&-\frac{\sqrt{2} g \sqrt{\cos 2 \theta_W}}{\cos \theta_W} v_R ,
\nonumber \\
    E&=&\frac{g \sqrt{\cos 2 \theta_W}}{\sqrt{2}\cos \theta_W} \kappa_d ,
\nonumber \\
    F&=& \frac{e }{\sqrt{2}}(\kappa_u+\kappa_d),
\end{eqnarray}
The unitary matrix $N$ would be replaced by a new matrix:
$N^{\prime}$ given by:
\begin{eqnarray}
N_{j1}^{\prime}&=&N_{j1}  \sin \theta_W + N_{j2} \sin \theta_W+ N_{j3}
\sqrt{\cos 2 \theta_W} \nonumber \\
N_{j2}^{\prime}&=&N_{j1} \cos \theta_W  - N_{j2} \sin \theta_W \tan
\theta_W - N_{j3} \sqrt{\cos 2 \theta_W} \tan \theta_W \nonumber \\
N_{j3}^{\prime}&=&N_{j2} \frac{\sqrt{\cos 2 \theta_W}}{\cos \theta_W} -
N_{j3} \tan \theta_W \nonumber \\
N_{jk}^{\prime}&=&N_{jk}, \ (k=4, 5, \ldots 9).
\end{eqnarray}
Similarly $N$ can be expressed in term of $N^{\prime}$ by:
\begin{eqnarray}
N_{j1}&=&N_{j1}^{\prime} \sin \theta_W + N_{j2}^{\prime} \cos \theta_W
\nonumber \\
N_{j2}&=&N_{j1}^{\prime} \sin \theta_W - N_{j2}^{\prime} \sin \theta_W
\tan \theta_W - N_{j3}^{\prime} \frac{\sqrt{\cos 2 \theta_W}}{ \cos
\theta_W} \nonumber \\
N_{j3}&=&N_{j1}^{\prime} \sqrt{\cos 2 \theta_W} - N_{j2}^{\prime} \tan
\theta_W \sqrt{\cos 2 \theta_W} \nonumber - N_{j3} \tan \theta_W \\
N_{jk}&=&N_{jk}^{\prime}, \ (k=4, 5, \ldots 9).
\end{eqnarray}

\subsection{Squarks}

In the interaction basis, $(\tilde{q}_L^{i}, \tilde{q}_R^{i})$, the
squared-mass matrix for a squark of flavor $f$ has the form:
\begin{equation}
{\cal M}_f^2= \left( \begin{array}{cc}
                             m_{f,LL}^2+F_{f,LL}+D_{f,LL} &
(m_{f,LR}^2)+F_{f,LR} \\
                             (m_{f,LR}^2)^{\dagger}+F_{f,RL} &
m_{f,RR}^2+F_{f,RR}+D_{f,RR}
                        \end{array}
                 \right).
\end{equation}
The F-terms are diagonal in the flavor space, $F_{f,LL, RR}=m_f^2$,
$(F_{d \ LR})_{ij}=-\mu
(m_{d_i} \tan \beta ) {\bf 1}_{ij}$,  $(F_{u \ LR})_{ij}=-\mu
(m_{u_i} \cot \beta ) {\bf 1}_{ij}$.
   The D-terms is also flavor-diagonal:
\begin{eqnarray}
D_{f \ LL}&=&M_Z^2 \cos 2 \beta (T_f^3-Q_f \sin^2 \theta_W) {\bf 1} 
\nonumber \\
D_{f \ RR}&=&M_Z^2 \cos 2 \beta Q_f \sin^2 \theta_W {\bf 1}
\end{eqnarray}
The term $(m_{f,LL,RR}^2)_{ij}=m^2_{\tilde{Q}_{L,R}} \delta_{ij}$, $m_{f
\ LR}^2=A_f^* m_f$.
To reduce the number of free parameters, we consider the parameters
to be universal,
with:
$(m^2_{\tilde{Q}_{L,R}})_{ij}=m_0^2 \delta_{ij}$, $A_{d,ij}=A
\delta_{ij}$ and $A_{u,ij}=A \delta_{ij}$.

The squared-mass matrix for U-type squarks reduces to:
\begin{equation}
{\cal M}_{U_k}^2= \left( \begin{array}{cc}
                             m_0^2+M_Z^2(T_u^3-Q_u \sin^2 \theta_W) \cos 2
\beta & m_{u_k} (A-\mu \cot \beta) \\
                            m_{u_k} (A-\mu \cot \beta) &
m_0^2+M_Z^2 Q_u \sin^2 \theta_W \cos 2 \beta
                        \end{array}
                 \right).
\end{equation}
with the diagonal F-terms absorbed into the $m_0^2$. For D-type squarks:
\begin{equation}
{\cal M}_{D_k}^2= \left( \begin{array}{cc}
                             m_0^2+M_Z^2(T_d^3-Q_d \sin^2 \theta_W) \cos 2
\beta & m_{d_k} (A-\mu \tan \beta)\\
                            m_{d_k} (A-\mu \tan \beta) &
m_0^2+M_Z^2 Q_d \sin^2 \theta_W \cos2 \beta
                        \end{array}
                 \right).
\end{equation}
The corresponding mass eigenstates are defined as:
\begin{equation}
{\tilde q}_{L,R}=\Gamma^{\dagger}_{Q \ L,R} \tilde{q}
\end{equation}
where $\Gamma^{\dagger}_{Q \ L,R}$ are 6$\times$ 3 mixing matrices.
In the universal case, there is no intergenerational mixings for
squarks and the only source of flavor mixing comes from the CKM
matrix. We will analyse this
case first. Next we will look at the case in which mixing in the squark
sector is permitted and consider
the effect of intergenerational mixings on the rate of the process $b 
\rightarrow s \gamma$.
As it is generally done
in the mass insertion approximation method \cite{MI}, where the
off-diagonal
squark mass matrix
elements are assumed to be small and
their higher orders can be neglected, we use normalized parameters:
\begin{eqnarray}
\label{massins}
\delta_{d,LL,ij}&=&\frac{(m^2_{d,LL})_{ij}}{m_0^2},~~~
\delta_{d,RR,ij}=\frac{(m^2_{d,RR})_{ij}}{m_0^2}, \nonumber \\
\delta_{d,LR,ij}&=&\frac{(m^2_{d,LR})_{ij}}{m_0^2},~~~
\delta_{d,RL,ij}=\frac{(m^2_{d,RL})_{ij}}{m_0^2},
\end{eqnarray}
In our analysis, we use the mass eigenstate formalism, which is valid
no matter how large
the intergenerational mixings are. We assume significant mixing between
the second and third
generations in the down-squarks mass matrix only.

\section{Supersymmetric contributions to $b \rightarrow s \gamma$}

\begin{figure}
\centerline{ \epsfysize 1.0in
\rotatebox{360}{\epsfbox{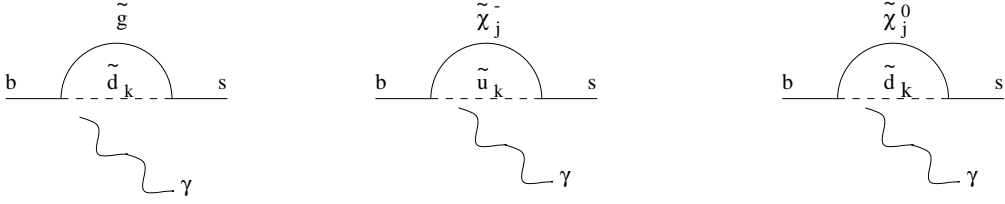}}  }
\caption{The Feynman diagrams contributing to the decay $b \rightarrow s
\gamma$. The outgoing photon line can be attached in all possible ways. }
\protect \label{feynmandiagrams}
\end{figure}

The low-energy effective Hamiltonian responsible for the B meson decay rates
at the scale
$\mu$ can be written as:
\begin{equation}
{\cal H}_{eff}=-\frac{4 G_F}{\sqrt{2}}K_{tb}K^*_{ts} \sum_i C_i(\mu)
Q_i(\mu).
\end{equation}
The operators relevant to the process $b \rightarrow s \gamma$ in LRSUSY
are:
\begin{eqnarray}
Q_7&=&\frac{e}{16 \pi^2}m_b(\mu) \bar{s} \sigma_{\mu \nu} P_R b F^{\mu
\nu},  \nonumber \\
Q_7^{\prime}&=&\frac{e}{16 \pi^2}m_b(\mu) \bar{s} \sigma_{\mu \nu} P_L
b F^{\mu \nu}, \nonumber \\
Q_8&=&\frac{g_s}{16 \pi^2}m_b(\mu) \bar{s} \sigma_{\mu \nu} G^{\mu
\nu}_a T^a P_R b ,  \nonumber \\
Q_8^{\prime}&=&\frac{g_s}{16 \pi^2}m_b(\mu) \bar{s} \sigma_{\mu \nu}
G^{\mu
\nu}_a T^a P_L b
\end{eqnarray}
and the Wilson coefficients $C_{7,8}^{\prime}$ are initially evaluated
at the
electroweak or soft supersymmetry breaking scale, then evolved down
to the scale $\mu$.
The Feynman diagrams contributing to this decay in LRSUSY are
illustrated in Fig.
\ref{feynmandiagrams}.

The matrix elements responsible for the $b \rightarrow s \gamma$
decay acquire the
following contributions from the supersymmetric sector of the model.
For $b_L$ decay:
\begin{equation}
M_{\gamma_R}=A_{\tilde{g}}^R+A_{\tilde{\chi}^-}^R+A_{\tilde{\chi}^0}^R
\end{equation}
with the gluino, chargino and neutralino contributions given by:
\begin{eqnarray}
A_{\tilde{g}}^R &=& - \frac{\pi \alpha_s}{\sqrt{2} G_F} Q_d C(R)
\sum_{k=1}^6
\frac{1}{m_{\tilde{d}_k}^2} \{ \Gamma_{DL}^{kb} \Gamma_{DL}^{*ks}
F_2(x_{\tilde{g} \tilde{d_k}})
-\frac{m_{\tilde{g}}}{m_b} \Gamma_{DR}^{kb} \Gamma_{DL}^{*ks}
F_4(x_{\tilde{g} \tilde{d_k}}) \} \\
A_{\tilde{\chi}^-}^R &=& - \frac{\pi \alpha_w}{\sqrt{2} G_F}
\sum_{j=1}^5 \sum_{k=1}^6
\frac{1}{m_{\tilde{u}_k}^2} \{
(G_{UL}^{jkb}-H_{UR}^{jkb})(G_{UL}^{*jks}-H_{UR}^{*jks})
[ F_1(x_{\tilde{\chi}_j \tilde{u}_k})+ Q_u F_2(x_{\tilde{\chi}_j
\tilde{u}_k})] \nonumber \\
    & & +\frac{m_{\tilde{\chi}_j}}{m_b} (G_{UR}^{jkb}-H_{UL}^{jkb})
(G_{UL}^{*jks}
-H_{UR}^{*jks}) [F_3 (x_{\tilde{\chi}_j \tilde{u}_k})+ Q_u
F_4(x_{\tilde{\chi}_j \tilde{u}_k})] \} \\
A_{\tilde{\chi}^0}^R &=& - \frac{\pi \alpha_w}{\sqrt{2} G_F} Q_d
\sum_{j=1}^9 \sum_{k=1}^6
\frac{1}{m_{\tilde{u}_k}^2} \{
(\sqrt{2}G_{0DL}^{jkb}-H_{0DR}^{jkb})(\sqrt{2}G_{0DL}^{*jks}-H_{0DR}^{*jks})
F_2(x_{\tilde{\chi}_j^0 \tilde{d}_k}) \nonumber \\
    & & +\frac{m_{\tilde{\chi}_j^0}}{m_b}
(\sqrt{2}G_{0DR}^{jkb}-H_{DL}^{jkb}) (\sqrt{2} G_{0DL}^{*jks}
-H_{0DR}^{*jks})  F_4(x_{\tilde{\chi}_j^0 \tilde{d}_k}) \}
\end{eqnarray}
and, for the decay of $b_R$:
\begin{equation}
M_{\gamma_L}=A_{\tilde{g}}^L+A_{\tilde{\chi}^-}^L+A_{\tilde{\chi}^0}^L
\end{equation}
again, with the following gluino, chargino and neutralino contributions:
\begin{eqnarray}
A_{\tilde{g}}^L &=& - \frac{\pi \alpha_s}{\sqrt{2} G_F} Q_d C(R)
\sum_{k=1}^6
\frac{1}{m_{\tilde{d}_k}^2} \{ \Gamma_{DR}^{kb} \Gamma_{DR}^{*ks}
F_2(x_{\tilde{g} \tilde{d_k}})
-\frac{m_{\tilde{g}}}{m_b} \Gamma_{DL}^{kb} \Gamma_{DR}^{*ks}
F_4(x_{\tilde{g} \tilde{d_k}}) \} \\
A_{\tilde{\chi}^-}^L &=& - \frac{\pi \alpha_w}{\sqrt{2} G_F}
\sum_{j=1}^5 \sum_{k=1}^6
\frac{1}{m_{\tilde{u}_k}^2} \{
(G_{UR}^{jkb}-H_{UL}^{jkb})(G_{UR}^{*jks}-H_{UL}^{*jks})
[ F_1(x_{\tilde{\chi}_j \tilde{u}_k})+ Q_u F_2(x_{\tilde{\chi}_j
\tilde{u}_k})] \nonumber \\
    & & +\frac{m_{\tilde{\chi}_j}}{m_b} (G_{UL}^{jkb}-H_{UR}^{jkb})
(G_{UR}^{*jks}
-H_{UL}^{*jks}) [F_3 (x_{\tilde{\chi}_j \tilde{u}_k})+ Q_u
F_4(x_{\tilde{\chi}_j \tilde{u}_k})] \} \\
A_{\tilde{\chi}^0}^L &=& - \frac{\pi \alpha_w}{\sqrt{2} G_F} Q_d
\sum_{j=1}^9 \sum_{k=1}^6
\frac{1}{m_{\tilde{u}_k}^2} \{
(\sqrt{2}G_{0DR}^{jkb}-H_{0DL}^{jkb})(\sqrt{2}G_{0DR}^{*jks}-
H_{0DL}^{*jks})
F_2(x_{\tilde{\chi}_j^0 \tilde{d}_k}) \nonumber \\
    & & +\frac{m_{\tilde{\chi}_j^0}}{m_b}
(\sqrt{2}G_{0DL}^{jkb}-H_{DR}^{jkb}) (\sqrt{2} G_{0DR}^{*jks}
-H_{0DL}^{*jks})  F_4(x_{\tilde{\chi}_j^0 \tilde{d}_k}) \}
\end{eqnarray}
where vertex mixing matrices $G$, $H$, $G_0$ and $H_0$ are defined in
the Appendix.
The convention
$x_{ab}=m_a^2/m_b^2$ is used. $C(R) =4/3$ is the quadratic Casimir
operator of the fundamental
representation of $SU(3)_C$.

In order to compare the results obtained with experimental branching
ratios, QCD
corrections must be taken into account.
We assume below the SM renormalization group evolution pattern;
supersymmetric estimates
exist for the gluino
contributions only \cite{bghw}. There is no mixing between left and
right-handed contributions.
\begin{equation}
A^{\gamma} (m_b)=\eta^{-16/23} \{ A^{\gamma} (M_W) + A^{\gamma}_0 [
\frac{116}{135} (\eta^{28/23}-1)
+\frac{58}{189}(\eta^{28/23}-1)] \},
\end{equation}
where $\eta=\alpha_s(m_b)/ \alpha_s(M_W)$ and $A^{\gamma}_0= \frac{\pi
\alpha_w }{2 \sqrt{2} G_F}
\frac{1}{M_W^2}$. We choose the renormalization scale to be $\mu=m_b=4.2$ GeV.

The inclusive decay width for the process $b \rightarrow s \gamma$ is
given by:
\begin{equation}
\Gamma(b \rightarrow s \gamma)=\frac{m_b^5 G_F^2 |K_{tb} K_{ts}^*|^2
\alpha}{32 \pi^4}
\left( \hat{M}_{\gamma L}^2+\hat{M}_{\gamma R}^2
\right),
\end{equation}
where the hat means evolving down to the decay scale $\mu=m_b$.
The branching ratio can be expressed as
\begin{equation}
BR (b\rightarrow s \gamma)= \frac{\Gamma (b\rightarrow s
\gamma)}{\Gamma_{SL}} BR_{SL},
\end{equation}
where the semileptonic branching ratio $BR_{SL}=BR(b \rightarrow ce
{\bar \nu})=(10.49
\pm 0.46)\%$ and:
\begin{equation}
\Gamma_{SL}=\frac{m_b^5 G_F^2 |K_{cb} |^2 }{192 \pi^3}g(z),
\end{equation}
where $z=m_c^2/m_b^2$ and $g(z)=1-8z+8z^3-z^4-12z^2 \mathrm{log}$$z$.

\section{Numerical results}

We are interested in analysing the case in which the supersymmetric
partners have
masses around the weak scale, so we will assume relatively light
superpartner masses. We diagonalize the neutralino and chargino mass matrices
numerically and we require in all calculations that the masses of 
gluinos, charginos,
neutralinos and squarks be above their experimental bounds. There are 
some extra
constraints in the
non-supersymmetric sector of the theory, requiring the FCNC Higgs
boson $\Phi_d$ to be heavy,
but no such constraints exist in the Higgsino sector \cite{pospelov}.
We choose the gluino
mass
$m_{\tilde g}=300$ GeV, and left-handed gaugino masses of $M_L=500$
GeV. The mass of the
lightest bottom squark will be in the $150-200$ GeV range. We include
in our numerical
estimates the non-supersymmetric LRM results, and for this we
constrain the lightest Higgs
mass to be
$115$ GeV \cite{higgs}. We analyze our results for low and moderate
values of $\tan \beta$,
although we  study the dependence of the branching ratio on $\tan 
\beta$. We also investigate
the dependence of the
branching ratio on both positive and negative values of the $\mu$
parameter.

As a first step, we assume the only source of flavor violation to
come from the CKM
matrix. This scenario is related to the minimal flavor violation
scenario in
supergravity. This restricted possibility of flavor violation will 
set important constraints
on the parameter space of LRSUSY.

We then allow, in the second stage of our investigation, for new 
sources of flavor
violation coming from the
soft breaking terms. In MSSM, this scenario is known as the
unconstrained MSSM and there
the gluino contribution dominates. This is not so in LRSUSY, where the
chargino
contribution is important for low to intermediate values of $M_R$, 
the right-handed
gaugino mass
parameter. We restrict all allowable LL, LR, RL and RR sflavor
mixing, assuming them to
be dominated by mixings between the second and third squark family.

We now proceed to discuss both these scenarios in turn.

\subsection{The constrained LRSUSY}

By the constrained LRSUSY model, we mean the scenario
in which the only source
of flavor violation comes from the quark sector, through the CKM
matrix, which we assume
to be the same for both the left and right handed sectors.

Before any meaningful numerical results be obtained, explicit values for
the
parameters in the model must be specified. There are many parameters in
the model
such that it is hard, if not impossible, to get an illustrative
presentation of
calculation results. If LRSUSY is embedded in a supersymmetric grand
unification
theory such as $SO(\mathrm{10})$,  there exist some relationships among
the parameters
at the unification scale $M_{GUT}$. We can generally choose specific
values for much less
parameters at mass scales $\mu= M_{GUT}$, then use renormalization 
group equations to run
them down to the low energy scale
which is relevant to
phenomenology. But, for maintaining both simplicity and generality, 
we can present an
analysis in which LRSUSY is not embedded into another group. Then we 
can choose all
parameters as independently free parameters, with the numerical 
results confronting with
experiments directly.

To make the results tractable, we assume all trilinear scalar couplings
in the soft
supersymmetry breaking Lagrangian as $A_{ij}=A \delta_{ij}$ and
$\mu_{ij}=\mu \delta_{ij}$, and we fix $A$ to be $50$ GeV in all the
analysis. We also set a common mass parameter for all the squarks
$M_{0UL}=M_{0UR}=
M_{0DL}=M_{0DR}=m_0$. The general range of these
parameters is as discussed in the previous paragraphs.
We also take $K^L_{CKM}=K^R_{CKM}$. This choice is conservative, and
much larger values
of mixing matrix elements are allowed in scenarios that attempt to
explain the decay
properties of the $b$ quark as being saturated by the right-handed $b$
\cite{gronau}. Our
choice does not favor one handedness over the other, and has the
added advantage that no
new mixing angles are introduced in the quark matrices.

\begin{figure}
\centerline{ \epsfysize 4.0in \rotatebox{270}{\epsfbox{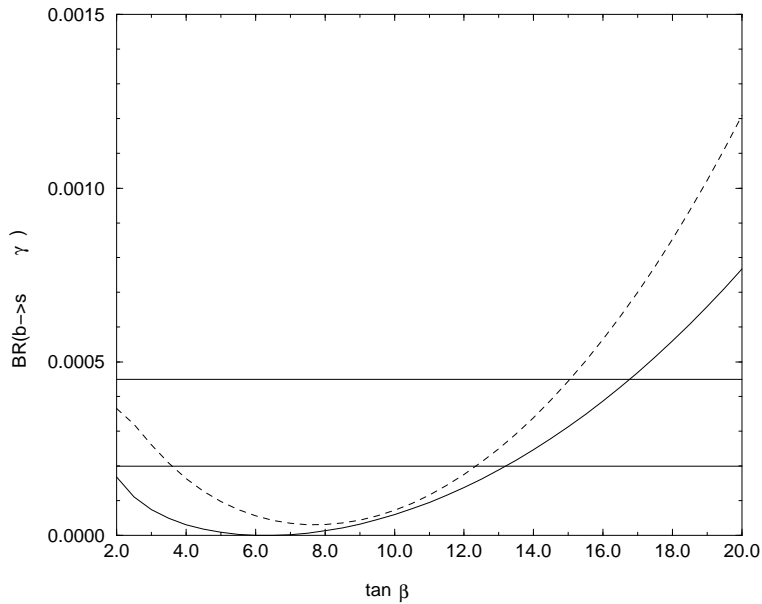}}  }
\caption{Supersymmetric contributions to BR($b \rightarrow s
\gamma$) as a function of $\tan \beta$, obtained when
$m_{\tilde{g}}=300$ GeV,
$\mu=100$ GeV, $M_L=M_R=500$ GeV and $m_0=200$ GeV. The full
contributions is also shown(dashed). The range of acceptable values of
branching ratios is given.}
\protect \label{figbeta}
\end{figure}

We investigate first the dependence of the branching ratio on the
values of $\tan
\beta$ in Fig. \ref{figbeta}. The solid line represents the
supersymmetric
contribution, the dashed the total contribution (including the SM and
LRM). All the graphs
show the experimental bounds as horizontal lines in the figures. The
universal squark mass is
around 200 GeV, and the chargino and neutralino are light. The choice of
parameters puts
stringent restrictions on the allowed values for $\tan
\beta$: either very low ($\tan \beta =$2-4), or intermediate in a very
small range ($\tan
\beta =$12-14) values are allowed.
As $\tan \beta$ becomes large, the branching
ration is almost linearly proportional to $\tan \beta$. For larger 
values of $\tan \beta$ the
branching ratio will exceed the acceptable range easily.
In our analysis, larger values of $\tan \beta$ are
allowed only for a
heavier supersymmetric mass spectra. For $sign(\mu)<0$, the range of 
acceptable intermediate
values of $\tan \beta $ increases. For example, if $\mu=-100$ GeV, 
the larger range $\tan
\beta =22-33$ is allowed.
We note that constraints on supersymmetry with large $\tan \beta$ were studied
in Ref. \cite{beta} where $\tan \beta$-enhanced chargino and charged
Higgs contributions were resummed to all order in perturbation theory.

\begin{figure}
\centerline{ \epsfysize 4.0in \rotatebox{270}{\epsfbox{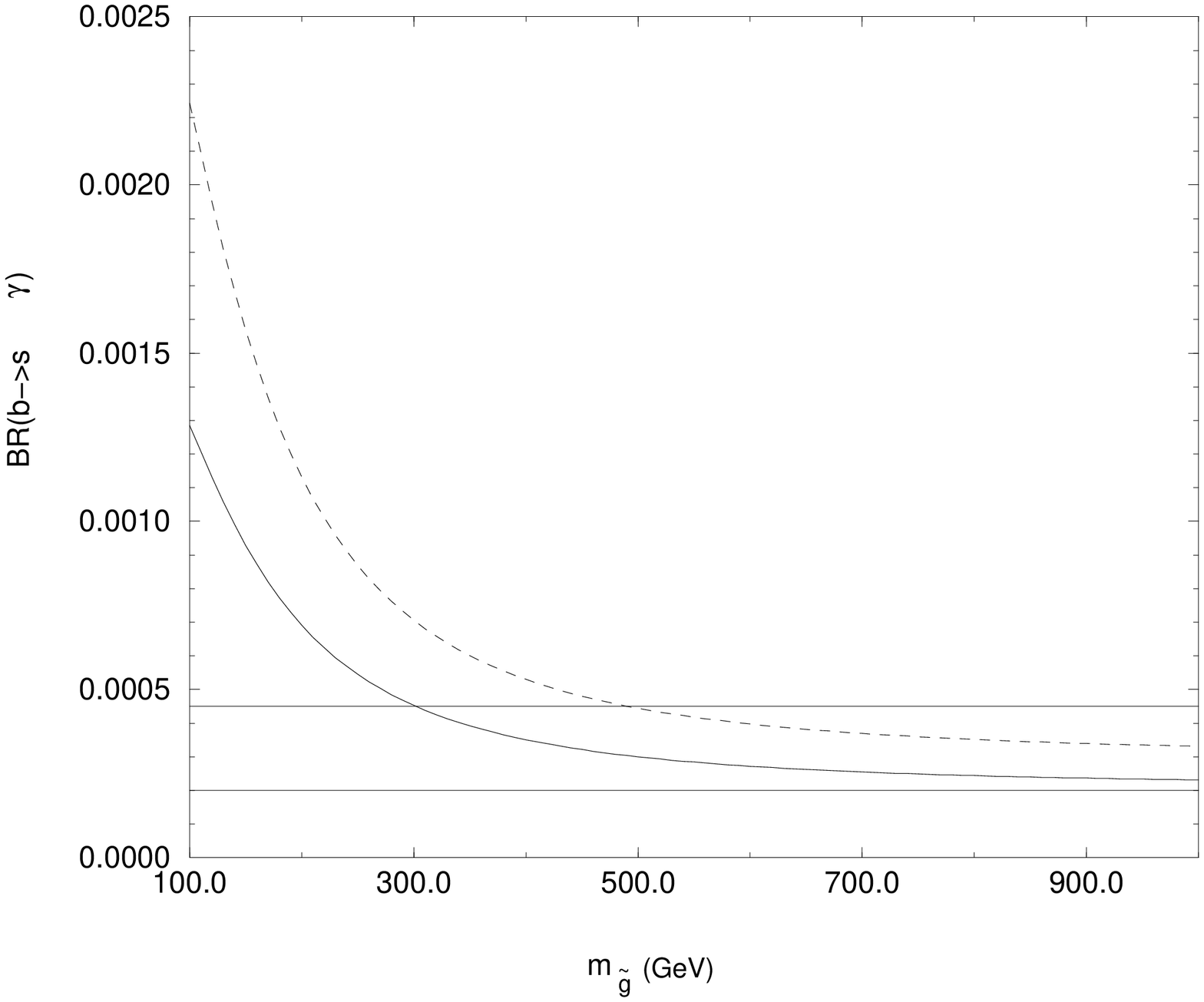}}  }
\caption{Supersymmetric contributions to BR($b \rightarrow s
\gamma$) as a function of the mass of the gluino $m_{\tilde{g}}$,
obtained when $\tan \beta =5$,
$\mu=100$ GeV, $M_L=M_R=500$ GeV and $m_0=100$ GeV. The full
contributions is also shown(dashed). The range of acceptable values of
branching ratios is given.}
\protect \label{figmgluino}
\end{figure}

As a general feature of the LRSUSY branching ratio, in a large region 
of parameter space, the
chargino contribution is
comparable to the gluino, while
the neutralino contribution is always smaller.
We investigate the dependence of
the branching ratio on the gluino mass, for a light squark scenario.
The chargino and
neutralino masses are light and $\mu/M_{L,R} \sim {\cal O}(1)$, a scenario
favored by recent analyses of the
anomalous magnetic moment of the muon \cite{g-2}. We present the 
results in Fig.
\ref{figmgluino}. The gluino
is constrained to be heavier than 300 GeV, albeit for a very light
supersymmetric
spectrum, close to experimental limits.

\begin{figure}
\centerline{ \epsfysize 4.0in \rotatebox{270}{\epsfbox{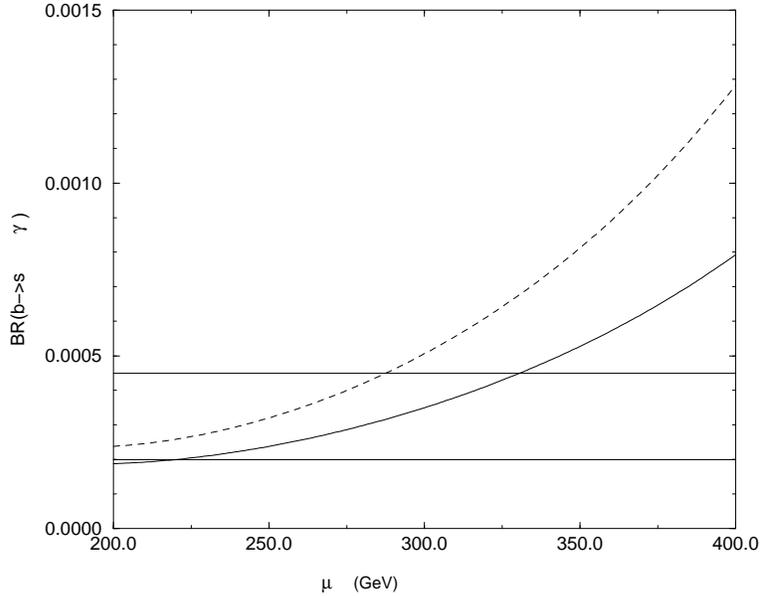}}  }
\caption{Supersymmetric contributions to BR($b \rightarrow s
\gamma$) as a function of $\mu$, obtained when $\tan \beta =5$,
$m_{\tilde{g}}=300$ GeV,
$M_L=M_R=500$ GeV and $m_0=100$ GeV. The full contributions is also
shown(dashed). The range of acceptable values of branching ratios is
given.}
\protect \label{figmu1}
\end{figure}

In the next two figures, we investigate the dependence of the
branching ratio of $b
\rightarrow s \gamma$ to the sign and magnitude of the Higgsino
mixing parameter $\mu$.
There have been indications that the new accurate measurement of the
anomalous magnetic
moment of the muon restrict the $\mu$ parameter to be positive, while $b
\rightarrow s \gamma$ favors a negative sign. For a light
squark-gaugino scenario, and
low $\tan \beta$, the bound on $b \rightarrow s \gamma$ is satisfied
for either sign of
the $\mu$ parameter. In Fig. \ref{figmu1} one could see that a
restricted region of
intermediate values for $\mu$, with sign$(\mu)>0$ is allowed by the
experimental
constraints on $b \rightarrow s \gamma$, in the 225-325 GeV region.
The parameter space
is less restrictive for $\mu$ negative, to -175 GeV. Note that for
$\mu \rightarrow 0$
the branching ratio drops outside the allowed range. This phenomenon occurs
because the mixing term
obtained from flipping chirality on the gaugino leg decouples. We
reject such small vaules of
the
$\mu$ parameter because the chargino and neutralino masses are smaller
than the
existing experimental bounds.

\begin{figure}
\centerline{ \epsfysize 4.0in \rotatebox{270}{\epsfbox{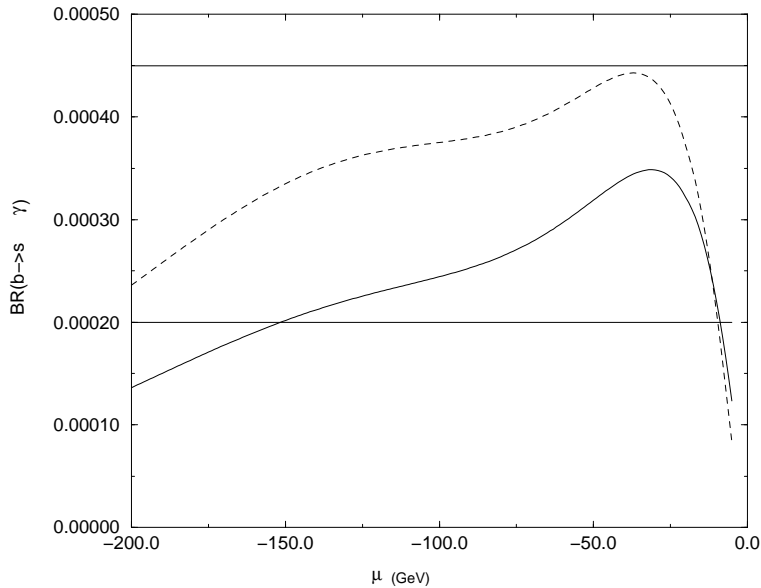}}}
\caption{Supersymmetric contributions to BR($b \rightarrow s
\gamma$) as a function of $\mu$, obtained when $\tan \beta =5$,
$m_{\tilde{g}}=300$ GeV,
$M_L=M_R=500$ GeV and $m_0=150$ GeV. The full contributions is also
shown(dashed). The range of acceptable values of branching ratios is
given.}
\protect \label{figmunegative}
\end{figure}

\begin{figure}
\centerline{ \epsfysize 4.0in \rotatebox{270}{\epsfbox{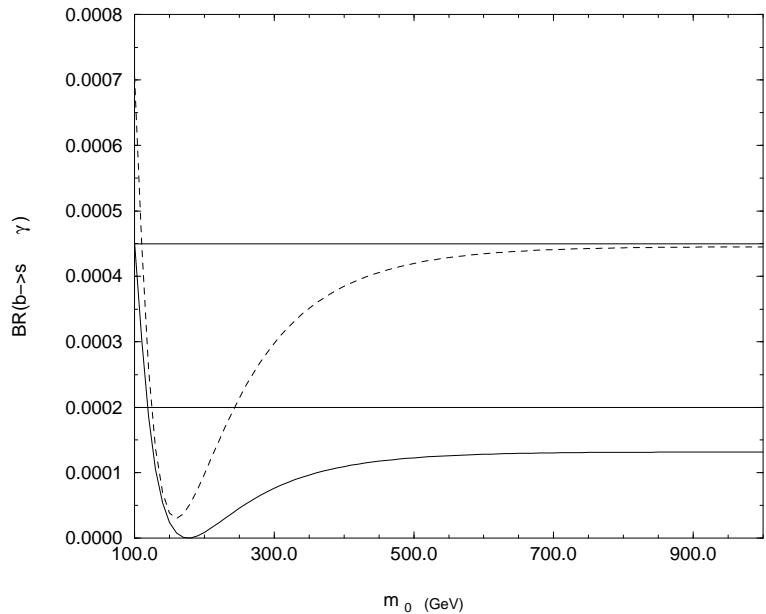}}  }
\caption{Supersymmetric contributions to BR($b \rightarrow s
\gamma$) as a function of $m_0$, obtained when $\tan \beta =5$,
$m_{\tilde{g}}=300$ GeV,
$M_L=M_R=500$ GeV and $\mu=100$ GeV. The full contributions is also
shown(dashed). The range of acceptable values of branching ratios is
given.}
\protect \label{figm0}
\end{figure}

The branching ratio for $b \rightarrow s \gamma$ is sensitive to the
universal scalar
mass $m_0$ in the region of small masses only. For $m_0 \ge 400$ GeV,
the branching
ratio reaches its QCD-corrected value and is stable against further
variations in the
scalar mass. In this scenario, the neutralinos and gluinos are light
and $\tan \beta=5$.
This situation is not unlike the dependence of the SM contribution on
the $t$ quark mass
\cite{smbsg, SUSY}.  This dependence is shown in Fig. \ref{figm0}.

\begin{figure}
\centerline{ \epsfysize 4.0in \rotatebox{270}{\epsfbox{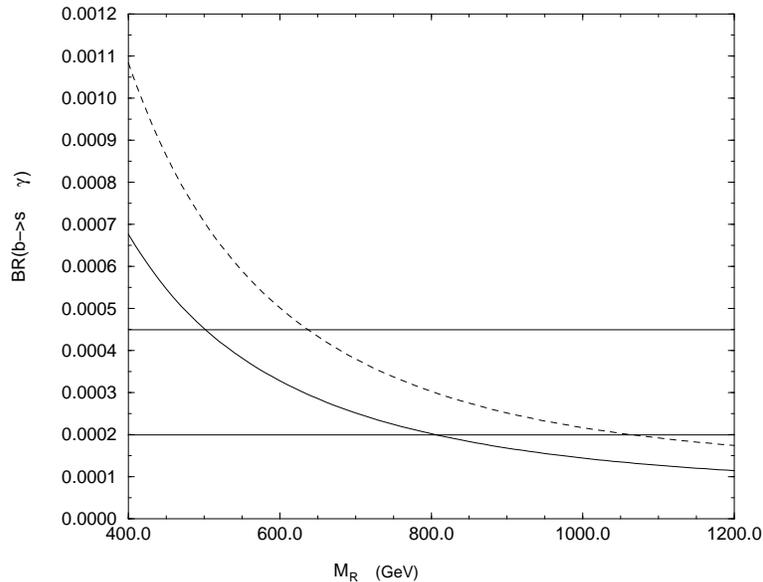}}  }
\caption{Supersymmetric contributions to BR($b \rightarrow s
\gamma$) as a function of $M_R$, obtained when $\tan \beta =5$,
$m_{\tilde{g}}=300$ GeV, $m_0=100$ GeV
$\mu=100$ GeV and $M_L=M_R$ is assumed. The full contributions is also
shown(dashed). The range of acceptable values of branching ratios is
given.}
\protect \label{figMR}
\end{figure}

In all the previous figures we set the left and right handed gaugino
masses to the same
value. This allowed a large contribution to the decay ratio of $b
\rightarrow s \gamma$ to come from the right-handed sector. We
investigate in
Fig. \ref{figMR} the dependence of the branching ratio on the
gaugino mass. As opposed
to the right-handed gauge sector, the restriction on the right handed
gaugino scale is not as
severe. There exist scenarios in which the right handed symmetry is
broken at the same
scale as supersymmetry; we expect in those cases to have approximately
$M_L=M_R$ \cite{kai}. For light squarks, Higgsinos and gluinos, the
gaugino mass must be
heavy, in the 600-800 GeV range.

\subsection{The unconstrained LRSUSY}

When supersymmetry is softly broken, there is no reason to expect
that the soft parameters
would be flavor blind, or that they would violate flavor in the same way as
in the SM.
Yukawa couplings generally form a matrix in the generation space, and
the off-diagonal elements will lead naturally to flavor changing
radiative decays.
Neutrino oscillations, in particular, indicate strong flavor
mixing
between the second and
third neutrino generations, and various analyses have been carried out
assuming the same for the
charged sleptons. In the quark/squark sector, the kaon system
strongly limits mixings
between the first and the second generations; but constraints for the
third generation are
much weaker, and expected to come from $b \rightarrow s \gamma$. The
unconstrained LRSUSY
model, similar to the unconstrained MSSM, allows for new sources of
flavor violation between
the second and third families only, both chirality conserving (LL and
RR) and chirality
flipping (LR and RL). We will assume that intergenerational mixing
occurs in the down squark
mass matrix only and that the up type squark mass matrix is diagonal.

With the definition of the mass insertion as in Eq. (\ref{massins}),
we can investigate the
effect of intergenerational mixing on the $b \rightarrow s \gamma$
decays. In the MSSM, the
branching ratio is dominated in this case by the gluino diagram, in
particular by the
chirality flip part of the gluino contribution, due to the
$\alpha_s/\alpha$ and $m_{\tilde
g}/m_b$ enhancements, respectively. In this case only the gluino
scenario is analysed in the
MSSM, and found to be dominated by $\delta_{23}^{RL}$ \cite{bghw}.
In LRSUSY, the situation
is different: the chargino graph contribution is comparable to the
gluino for a large range
of gaugino masses.

\begin{figure}
\centerline{ \epsfysize 4.0in \rotatebox{270}{\epsfbox{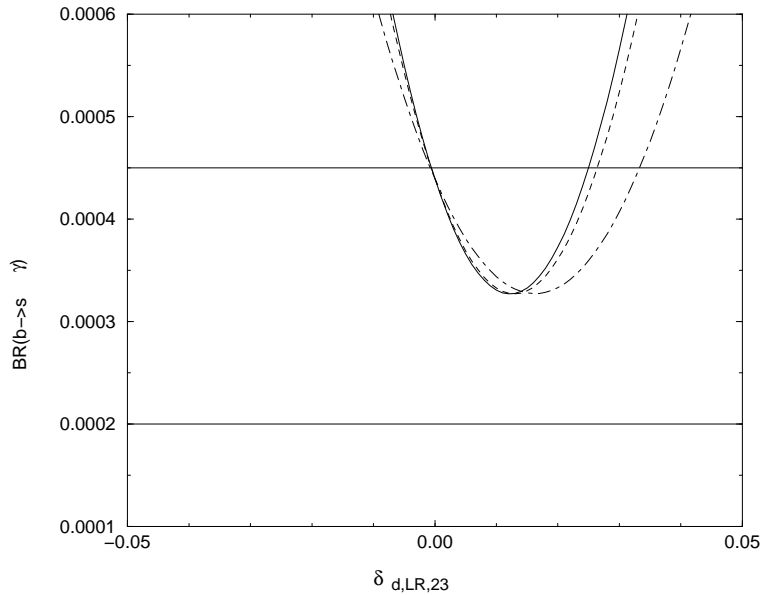}}}
\caption{Dependence of BR($b \rightarrow s
\gamma$) on $\delta_{d,LR,23}$, obtained when $\tan \beta =5$,
$\mu=500$ GeV and $M_L=M_R=500$ GeV. The different lines correspond to
different values
of $x=m^2_{\tilde{g}}/m^2_{0}$, 0.16(solid), 0.64(dashed) and
1.44(dot-dashed).
$m_0$ is fixed to be $500$ GeV.
The range of acceptable values of branching ratios is given.}
\protect \label{figdeltadLR23}
\end{figure}

We keep our analysis general, but to
show our results, we select only one possible source of flavor
violation in the squark
sector at a time, and assume the others vanish. All diagonal entries
in the squark
mass matrix are set equal and we study the branching ratio as a
function of their common
value $m_0^2$ and the relevant off-diagonal element. In Fig.
\ref{figdeltadLR23} we show the
dependence of $b \rightarrow s \gamma$ as a function of $\delta_{d,
LR,23}$ when this is the
only source of flavor violation. The horizontal lines represent the
range of values allowed
experimentally for the branching ratio. The ratio is plotted as a
function of different
values for the ratio $x=m^2_{\tilde{g}}/m^2_{0}$. Fixing $m_0=500$
GeV, this corresponds to
gluino masses of 200 GeV, 400 GeV and 600 GeV respectively. Negative
values of $\delta_{d,
LR,23}$ are more constrained than positive values, but in any case
$\delta_{d, LR,23} \le
4\%$.
This flavor violating parameter is strongly constrained because through the
$\delta_{d,LR,23}$ term, the helicity flip needed for $b \rightarrow 
s \gamma$ can be
realized in the exchange particle loop.
Comparison with the MSSM \cite{bghw,kane} shows that this
parameter is more
constrained in LRSUSY, but only slightly.

\begin{figure}
\centerline{ \epsfysize 4.0in \rotatebox{270}{\epsfbox{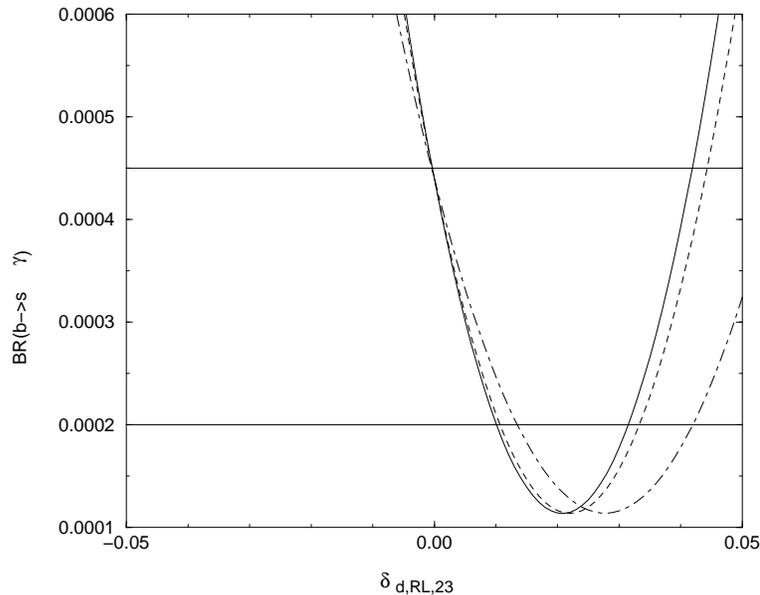}}}

\caption{Dependence of BR($b \rightarrow s
\gamma$) on $\delta_{d,RL,23}$, obtained when $\tan \beta =5$,
$\mu=500$ GeV and $M_L=M_R=500$ GeV. The different lines correspond to
different values
of $x=m^2_{\tilde{g}}/m^2_{0}$, 0.16(solid), 0.64(dashed) and
1.44(dot-dashed).
$m_0$ is fixed to be $500$ GeV.
The range of acceptable values of branching ratios is given.}
\protect \label{figdeltadRL23}
\end{figure}

The situation is very different when the only source of flavor violation
is
$\delta_{d, RL,23}$, as shown in Fig. \ref{figdeltadRL23}. MSSM results for
$b
\rightarrow s \gamma$ are
symmetric around
$\delta_{d, RL,23}=0$ and the experimental
bounds are satisfied for any small values of $\delta_{d, RL,23}$ . In
LRSUSY, practically no
negative values of $\delta_{d, RL,23}$ satisfy the bounds, and this
flavor violating
parameter is less restricted than $\delta_{d, LR,23}$ for the same
values of the squark and
gluino masses.

\begin{figure}
\centerline{ \epsfysize 4.0in \rotatebox{270}{\epsfbox{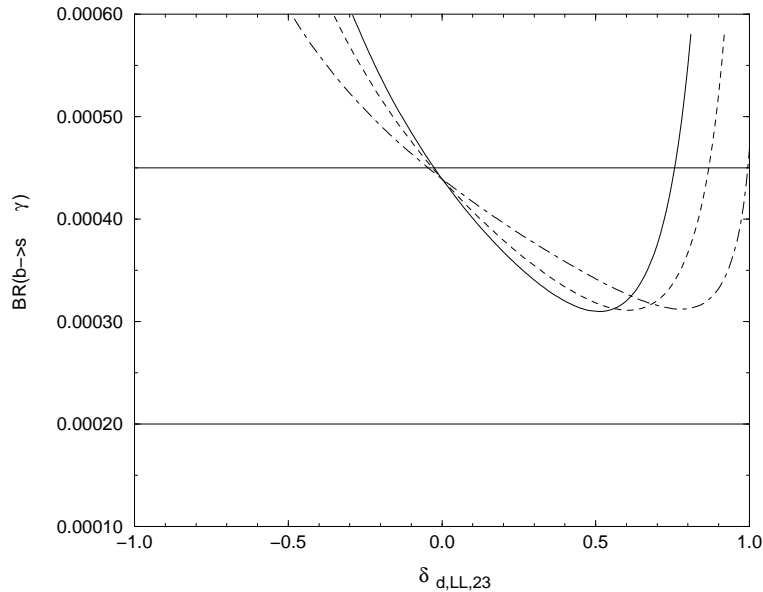}}}
\caption{Dependence of BR($b \rightarrow s
\gamma$) on $\delta_{d,LL,23}$, obtained when $\tan \beta =5$,
$\mu=400$ GeV and $M_L=M_R=500$ GeV. The different lines correspond to
different values
of $x=m^2_{\tilde{g}}/m^2_{0}$, 0.16(solid), 0.64(dashed) and
1.44(dot-dashed).
$m_0$ is fixed to be $500$ GeV.
The range of acceptable values of branching ratios is given.}
\protect \label{figdeltadLL23}
\end{figure}

\begin{figure}
\centerline{ \epsfysize 4.0in \rotatebox{270}{\epsfbox{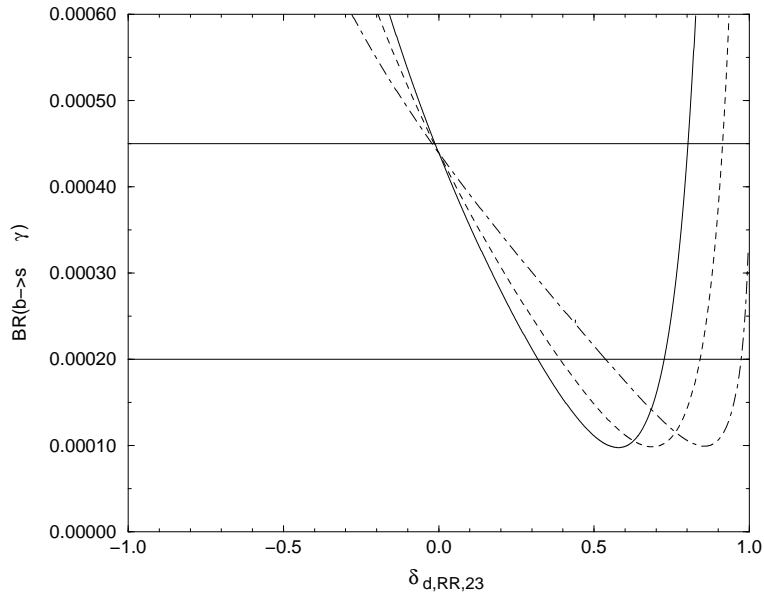}}}

\caption{Dependence of BR($b \rightarrow s
\gamma$) on $\delta_{d,RR,23}$, obtained when $\tan \beta =5$,
$\mu=500$ GeV and $M_L=M_R=500$ GeV. The different lines correspond to
different values
of $x=m^2_{\tilde{g}}/m^2_{0}$, 0.16(solid), 0.64(dashed) and
1.44(dot-dashed).
$m_0$ is fixed to be $500$ GeV.
The range of acceptable values of branching ratios is given.}
\protect \label{figdeltadRR23}
\end{figure}

In Fig. \ref{figdeltadLL23} and Fig. \ref{figdeltadRR23} we plot the
dependence of the
branching ratio of $b \rightarrow s \gamma$ on the chirality conserving
mixings $\delta_{d,LL,23}$ and
$\delta_{d,RR,23}$ respectively, with the proviso that these are the
only off-diagonal
matrix elements in the squark mass matrix squared. Although the
restriction is not as
pronounced as the one for chirality flipping parameters, nonetheless
the parameter
$\delta_{d,LL,23}$ is more restricted if it is negative (to 50\%)
than if positive (where almost
all values allowed for large gluino masses), quite different than in
the MSSM, where values
centered around
$\delta_{d,LL,23}=0$ were favored \cite{bghw}. The same is true for
the parameter
$\delta_{d,RR,23}$ which in MSSM was restricted slightly only for
  $\pm$ 100\% values, but in
LRSUSY regions of restrictions are centered around 50\% , and
increasing with gluino mass
for fixed scalar mass; again, any negative values are ruled out by
the experimental bounds
in the parameter region considered.

In Ref. \cite{Gabbiani}, a detailed analysis of FCNC and CP
constraints on these parameters
was presented. For the decay $b \rightarrow s \gamma$, only poor
constraints on $\delta_{d,LL, 23}$
existed, while $\delta_{d, LR, 23}$ was found to be constrained
strongly. This is compatible with our
analysis even though only the gluino-mediated contribution to the decay was
considered there.

\section{Conclusions}

We have presented a detailed and complete analysis of all one-loop
contributions to the
branching ratio of $b \rightarrow s \gamma$ in the LRSUSY model. We
analysed separately the
case in which the only source of flavor violation comes from the
quark sector (CKM matrix).
We refer to that case as the constrained LRSUSY model, in analogy
with MSSM. If we allow
for soft-supersymmetry intergenerational mixing in the squark sector,
new sources flavor
violation can occur; we refer to that case as the unconstrained
LRSUSY and we analyse it and
compare it to MSSM under similar conditions.

The model contains too many parameters to allow for a precise
restriction on any single one.
However as a general feature, some constraints arise for low squark
masses. In the
constrained LRSUSY case, for intermediate gluino-neutralino masses,
the $\mu$ parameter is
favored to be such that
$\mu/M_{L,R} \sim {\cal O}(1)$, and a larger region of parameter space 
satisfies the experimental
constraints for
$sign(\mu)<0$ than for
$sign(\mu)>0$. A small range of low or intermediate values of $\tan
\beta $ are allowed for
such a choice: for larger values of $\tan \beta $ the gaugino,
Higgsino and squark masses
must be higher. The branching ratio is relatively insensitive to values
of squark masses above
500 GeV, where the branching ratio becomes equal to its QCD-corrected
value. For a light
neutralino-chargino scenario, the mass of the gluino must be $ \ge
300$ GeV. For a gluino mass
of order 300 GeV and very light squarks ($m_{\tilde t}=100$ GeV), the
left and/or right
gaugino masses must be in the 600-800 GeV range.

For the unconstrained LRSUSY model, assuming flavor mixing only
between the second and third
generation in the down squark mass mixing matrix, the branching ratio
is dominated by the
internal chirality flipping diagrams, as in MSSM. Here however, the
chargino graphs are
comparable to the gluino contributions. The model puts stricter
constrains on the chirality
flipping mass mixings
$\delta_{d,LR,23}$ and
$\delta_{d,RL,23}$ than the chirality conserving flavor mixing parameters
$\delta_{d,LL,23}$ and
$\delta_{d,RR,23}$. The difference between LRSUSY and MSSM is quite
striking in restrictions
on the chirality conserving
$\delta_{d,LL,23}$ and $\delta_{d,RR,23}$. As opposed to MSSM where
both negative and
positive values of these parameters are allowed, LRSUSY severely
restricts the range of the
negative values. This is understood as a consequence of the
left-right structure of the
gauge-gaugino sector. For $\delta_{d,RR,23}$, there seems to be a
small range of disallowed
values in a narrow range around 50\%. If the dominant sources of
flavor violation come from
chirality conserving sflavor mixing, the MSSM and LRSUSY allow for a
distinguishingly
different range of parameters.

\bigskip
\noindent {\bf Acknowledgements}

This work was funded by NSERC of Canada (SAP0105354).

\newpage

\begin{appendix}

\noindent {\Large {\bf Appendix}}

The relevant Feynman rules used in the calculation are listed in this
appendix.
The three vertices of gluino-quark-squark, chargino-quark-squark and
neutralino-quark-squark interactions are represented in Fig. 
\ref{figfeynmanrules}. From the
first graph:

\begin{figure}
\centerline{ \epsfysize 1.5in
\rotatebox{360}{\epsfbox{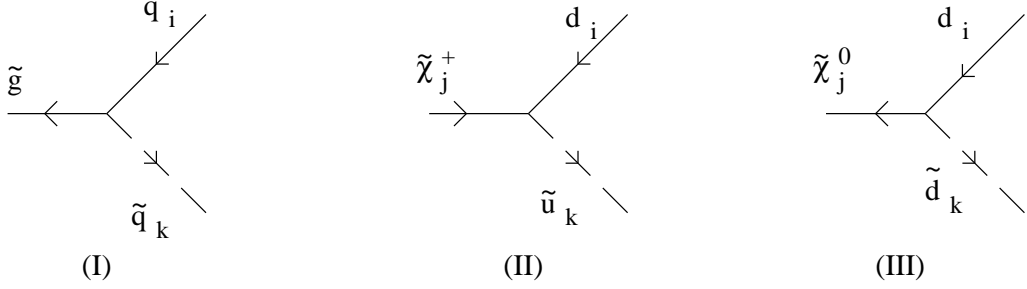}}  }
\caption{The Feynman rules used in the calculation. }
\protect \label{figfeynmanrules}
\end{figure}

\begin{equation}
(I)=-i g_s \sqrt{2} T^a_{\alpha \beta} (\Gamma_{QL}^{ki} P_L -
\Gamma_{QR}^{ki} P_R),
\end{equation}
where $P_{L,R}=(1 \pm \gamma_5)/2$, and $T^a$ are SU(3) color generators
normalized
to $Tr(T^aT^b)=\delta^{ab}/2$; and $\Gamma_{QL,R}$ are mixing matrices for
scalar quarks. From the second graph:
\begin{equation}
(II)=-i g C^{-1} [ (G_{UL}^{jki}-H_{UR}^{jki})P_L + (G_{UR}^{jki} -
H_{UL}^{jki}) P_R ],
\end{equation}
where $C$ is the charge conjugation operator (in spinor space) and the
chargino-quark-squark mixing martices $G$
and $H$ are defined as:
\begin{eqnarray}
G^{jki}_{UL} &=& V_{j1}^{\ast} (K_{CKM})_{il} (\Gamma_{UL})_{kl}
\nonumber \\
G^{jki}_{UR} &=& U_{j2} (K_{CKM})_{il} (\Gamma_{UR})_{kl} \nonumber \\
H_{UL}^{jki} &=& \frac{1}{\sqrt{2} m_W} ( \frac{m_{u_l}}{\sin \beta}
U_{j3}+
\frac{m_{d_l}}{\cos \beta} U_{j4} ) (K_{CKM})_{il} (\Gamma_{UL})_{kl}
\nonumber \\
H_{UR}^{jki} &=& \frac{1}{\sqrt{2} m_W} ( \frac{m_{u_l}}{\sin \beta}
V_{j3}^*+
\frac{m_{d_l}}{\cos \beta} V_{j4}^* ) (K_{CKM})_{il} (\Gamma_{UR})_{kl}.
\end{eqnarray}
Finally the contribution from the third graph is:
\begin{equation}
(III)=-i g [ (\sqrt{2} G_{0DL}^{jki}+H_{0DR}^{jki})P_L - (\sqrt{2}
G_{0DR}^{jki} - H_{0DL}^{jki}) P_R],
\end{equation}
where the neutralino-quark-squark mixing matrices $G_0$
and $H_0$ are defined as
\begin{eqnarray}
G^{jki}_{0DL} &=& [\sin \theta_W Q_d N^{\prime}_{j1} + \frac{1}{\cos
\theta_W} (T^3_{d}-Q_d \sin^2 \theta_W)
    N^{\prime}_{j2} \nonumber \\
& -&\frac{\sqrt{\cos 2 \theta_W }}{\cos \theta_W}
\frac{Q_u+Q_d}{2} N^{\prime}_{j3}
] (K_{CKM})_{il} (\Gamma_{DL})_{kl} \nonumber \\
G^{jki}_{0DR} &=& -[\sin \theta_W Q_d N^{\prime}_{j1} - \frac{Q_d \sin^2
\theta_W}{\cos \theta_W}
    N^{\prime}_{j2} \nonumber \\
&+&\frac{\sqrt{\cos 2 \theta_W }}{\cos \theta_W}
(T^3_{d}-Q_d \sin^2 \theta_W) N^{\prime}_{j3}
] (K_{CKM})_{il} (\Gamma_{DR})_{kl}  \nonumber \\
H_{0DL}^{jki} &=&  \frac{1}{\sqrt{2} m_W} ( \frac{m_{u_l}}{\sin \beta}
N^{\prime}_{j5}+
\frac{m_{d_l}}{\cos \beta} N^{\prime}_{j7}) (K_{CKM})_{il}
(\Gamma_{DL})_{kl} \nonumber \\
H_{0DR}^{jki} &=& \frac{1}{\sqrt{2} m_W} ( \frac{m_{u_l}}{\sin \beta}
N^{\prime *}_{j5}+
\frac{m_{d_l}}{\cos \beta} N^{\prime \ast}_{j7} ) (K_{CKM})_{il}
(\Gamma_{DR})_{kl}.
\end{eqnarray}

\end{appendix}

\def\oldprd#1#2#3{{\rm Phys. ~Rev. ~}{\bf D#1}, #3 (19#2)}
\def\newprd#1#2#3{{\rm Phys. ~Rev. ~}{\bf D#1}, #3 (20#2)}
\def\plb#1#2#3{{\rm Phys. ~Lett. ~}{\bf B#1}, #3 (#2)}
\def\newplb#1#2#3{{\rm Phys. ~Lett. ~}{\bf B#1}, #3 (20#2)}
\def\npb#1#2#3{{\rm Nucl. ~Phys. ~}{\bf B#1}, #3 (19#2)}
\def\newnpb#1#2#3{{\rm Nucl. ~Phys. ~}{\bf B#1}, #3 (20#2)}
\def\prl#1#2#3{{\rm Phys. ~Rev. ~Lett. ~}{\bf #1}, #3 (19#2)}
\def\prl20#1#2#3{{\rm Phys. ~Rev. ~Lett. ~}{\bf #1}, #3 (20#2)}
\def\rep19#1#2#3{{\rm Phys. ~Rep. ~}{\bf #1}, #3 (19#2)}
\def\rep20#1#2#3{{\rm Phys. ~Rep. ~}{\bf #1}, #3 (20#2)}
\def\epjc#1#2#3{{\rm Eur. ~Phys. J.~}{\bf C#1}, #3 (#2)}

\bibliographystyle{unsrt}

\end{document}